\documentclass[12pt,letterpaper,twoside,onecolumn,useAMS]{article}

\usepackage{tcolorbox}
\usepackage{natbib}
\usepackage[font={small,sf},labelfont={small,bf}]{caption}
\usepackage{wrapfig}
\usepackage{multicol}
\usepackage{url}
\usepackage{multirow}
\usepackage{floatrow}
\usepackage{bm}
\usepackage{textgreek}

\usepackage{times}
\usepackage{geometry}
\usepackage[utf8]{inputenc}
\usepackage{enumitem,amssymb}
\usepackage{ragged2e}
\newlist{thematic}{itemize}{8}
\setlist[thematic]{label=$\square$}
\usepackage{pifont}

\begin{document}
\begin{titlepage}
\raggedright
\huge
{\bf \centering{Astro2020 Science White Paper} \linebreak

A Survey of Hot Gas in the Universe
\linebreak}
\normalsize

\noindent \textbf{Thematic Areas:} \hspace*{60pt} $\square$ Planetary Systems \hspace*{10pt} $\square$ Star and Planet Formation \hspace*{20pt}\linebreak
$\square$ Formation and Evolution of Compact Objects \hspace*{31pt} $\square$ Cosmology and Fundamental Physics \linebreak
  $\square$  Stars and Stellar Evolution \hspace*{1pt} $\square$ Resolved Stellar Populations and their Environments \hspace*{40pt} \linebreak
  $\checkmark$    Galaxy Evolution   \hspace*{45pt} $\square$             Multi-Messenger Astronomy and Astrophysics \hspace*{65pt} \linebreak
  
Name:	Joel N. Bregman
 \linebreak						
Institution:  University of Michigan
 \linebreak
Email: jbregman@umich.edu
 \linebreak
 
\textbf{Co-authors:} Edmund Hodges-Kluck (University of Maryland/NASA GSFC), Benjamin D. Oppenheimer (University of Colorado), Laura Brenneman (Harvard-Smithsonian), Juna Kollmeier (Carnegie), Jiangtao Li (University of Michigan), Andrew Ptak (NASA GSFC), Randall Smith (Harvard-Smithsonian), Pasquale Temi (NASA Ames), Alexey Vikhlinin (Harvard-Smithsonian), Nastasha Wijers (Leiden)
  \linebreak

\textbf{Summary:} 
A large fraction of the baryons and most of the metals in the Universe are unaccounted for. They likely lie in extended galaxy halos, galaxy groups, and the cosmic web, and measuring their nature is essential to understanding galaxy formation. These environments have virial temperatures $\gtrsim$ 10$^{5.5}$~K, so the gas should be visible in X-rays. Here we show the breakthrough capabilities of grating spectrometers to 1) detect these reservoirs of hidden metals and mass, and 2) quantify hot gas flows, turbulence, and rotation around the Milky Way and external galaxies.  Grating spectrometers are essential instruments for future X-ray missions, and existing technologies provide 50--1500-fold higher throughput compared to current orbiting instruments.

\begin{figure}[!h]
\vspace{-0.0cm}
  \includegraphics[width=1\textwidth]{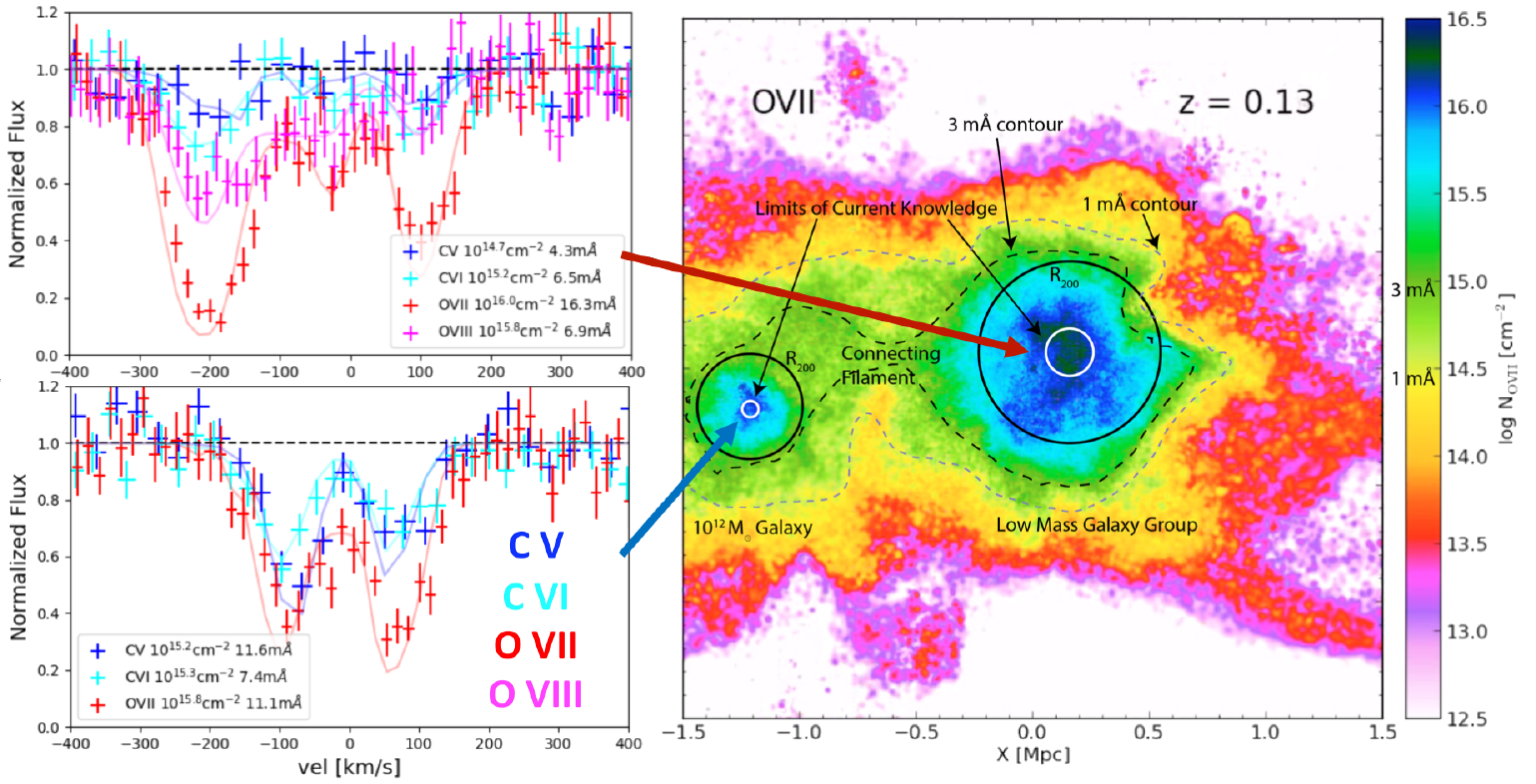}
\label{fig:mstar_color}
\vspace{-2.0cm}
\end{figure}

\end{titlepage}

\setcounter{page}{1}

\pagebreak

\noindent{\bf \large 1. Introduction and an Overview of X-Ray Spectroscopy}  
\medskip

The characteristic temperature of galaxies, galaxy clusters, stars, neutron stars, black hole accretion disks, and exploding objects occurs at T $>$ 0.5$\times$10$^{{\rm 6}}$ K.  At temperatures of 0.5$-$100$\times$10$^{{\rm 6}}$~K, diagnostic emission and absorption lines are from metals (i.e., O, C, Fe, Mg, Si, Ne), and most have energies in the X-ray band. Of these lines, the strong oxygen lines at $10^6$~K (O~{\sc vii} He$\alpha$ and O~{\sc viii} Ly$\alpha$, at 21.6\AA\ and 18.9\AA) are particularly important because of the oxygen abundance relative to other metals and variety of environments where they are detected. To measure line profiles and detect faint lines, the resolution must be close to the thermal width, which for the oxygen lines is $54 \times (T/10^6$ K)$^{1/2}$ km s$^{-1}$. This suggests a resolution of $R$=3,000-10,000.

Currently, the primary spectrographs are the Reflection Grating Spectrometer (RGS) on \textit{XMM-Newton} and the Low (and High) Energy Transmission Grating Spectrometers (L/HETGS) on the \textit{Chandra X-Ray Observatory}. The RGS has a resolution $R \sim 360$ at the key oxygen lines and a collecting area of about 45 cm$^{{\rm 2}}$. The same region is covered by the LETGS with a somewhat higher spectral resolution of $R$=400 but with lower collecting area, $\sim 7$ cm$^{2}$.  These instruments have led to major advances in a variety of fields, despite their sub-optimal resolution.

In the near future (2022), a micro-calorimeter with 5~eV resolution will be launched on {\it XRISM}, followed by the ESA \textit{Athena} X-IFU with 2~eV resolution \citep{barcons17}. Unlike gratings, which have a constant wavelength resolution ($\Delta \lambda$), calorimeters have constant energy resolution ($\Delta E$). Hence, \textit{XRISM} will resolve the thermal width ($\sim$200~km~s$^{-1}$) for the high ionization Fe lines detected above 6~keV, but below 1~keV (where most X-ray transitions lie) it cannot do so; the resolution is only 130 for O~{\sc viii} Ly$\alpha$, three times worse than the HETGS. Non-dispersive calorimeters will obtain spectra with better than CCD energy resolution for regions of diffuse, low surface brightness gas, but grating spectrometers are needed to access most of the information encoded in line profiles, as well as to detect weak absorption lines below 1~keV. 

The modest collecting area and sub-optimal resolution of current orbiting gratings and the inability of currently planned micro-calorimeters to resolve the thermal width for most X-ray transitions motivates a superior grating spectrometer. Current technology achieves $R$=3,000-10,000 and collecting areas of 300-4000 cm$^{{\rm 2}}$. This improves on the RGS and HETGS by orders of magnitude, opening many avenues of discovery. Some of the key ions and
lines, and the temperatures at which they are collisionally ionized are shown in Fig. 1. 
\medskip

\noindent{\bf \large 2. ``Missing'' Baryons and Metals Outside of Galaxies}  \medskip

One of the striking realizations of the past 30 years is that \textit{galaxies are missing most of their baryons and metals}  \citep{fuku98, mcgaugh10,dai2010, shull2012,shull2014}.  One obtains the dark matter mass from the rotation curve and the cosmic baryon fraction from CMB studies.  The observed baryonic mass comes from a sum of the stellar mass of stars (plus remnants), and the gas mass of the H~{\sc i}, H~{\sc ii} and H$_{{\rm 2}}$ disks of spirals or the hot atmospheres in early-type galaxies. These baryonic components only account for $\sim${}30\% of the cosmic baryon value for an L* galaxy, and this becomes more severe at lower L.

Theoretical models provide context for this problem and show that it is not just an accounting curiosity, but a crucial clue to understanding galaxy formation and evolution.  In simple spherical accretion models, potential energy is converted to thermal energy at an accretion shock, producing a massive hot halo at $\sim${}T$_{{\rm virial}}$ \citep{mo2010}.  Modern simulations show that accretion is more complex and asymmetric, with ensuing star formation leading to supernovae that heat their surroundings, driving winds from the galaxies.  Some gas is rendered unbound, while most remains bound to the CGM \citep[e.g.,][]{fielding17}, either as hot gas or warm (non-buoyant) clouds \citep{werk14,tumlinson17}. However, simulations differ greatly in the CGM mass, its temperature and metal distribution, the fraction of baryons beyond R$_{{\rm 200}}$, and the rate that gas cools onto the disk of the galaxy \citep{sch15,hop18,nel18a}.  The reason for these broad uncertainties is poor observational constraints on hot CGM.

A related issue is the \textit{missing metals problem}. 
The metals produced by a galaxy can be estimated from the observed stellar populations and the characteristic metal yield per supernova, and we find that only 1/4 of the metals are retained in the stars and
gas of the galaxy --  most of the metals were ejected through winds \citep{peep14} and presumably reside in the CGM.

A second missing baryon and missing metals problem exists \textit{for the Universe as a whole}, as 1/3-1/2 of the baryons are unaccounted for \citep{shull2012}, as are most of the metals. When calculating the total metal mass from the cosmic star-formation rate \citep{shull2014} or supernovae \citep{maoz17}, one predicts a cosmic metallicity $0.1-0.2 Z_{\odot}$. However, the metal budget from stars plus the CGM, as measured from UV absorption lines \citep[using the Cosmic Origins Spectrograph on {\it Hubble};][]{keeney17, pro17, wotta19}, only accounts for 20-40\% of this value.  A likely solution is that most metals lie in the currently  ``invisible'' hot diffuse gas, which must have a mean metallicity of 0.2-0.3 Solar to close the metal budget. This is good news for observers, as there should be plentiful ions to produce absorption lines.
\medskip

\noindent{\bf 2.1. A Blind Census of $10^{5.5}$-$10^{6.5}$~K Gas in the Universe}  \medskip

A prime observational goal is to execute an unbiased survey of hot gas in the Universe by measuring X-ray absorption systems for lines of sight toward distant AGNs.  This parallels the investigations using UV lines to characterize the neutral and warm ionized gas, and completes the census of gas in the Universe.  The expected temperatures and oscillator strengths imply that O~{\sc vii} and O~{\sc viii} will be the key ions, and one would obtain values for $\Omega$(O~{\sc vii}), $\Omega$(O~{\sc viii}), and other ions as a function of redshift. The few X-ray absorption lines detected so far \citep[e.g.][]{nicastro18,kovacs19} prove feasibility, but are inadequate for determining metallicities.

Such a survey connects directly to simulations of structure formation, which predict a range of equivalent width (EW) distributions (e.g. Fig. 2). Virialized systems have characteristic halo column densities of $\sim${}n$_{{\rm 200}}$R$_{{\rm 200}}$, or about 10$^{{\rm 19.5}}$ cm$^{{\rm -}{2}}$ for an L* galaxy, and this measure increases as M$_{h}^{1/3}$. Coupled with the metallicity needed to account for the missing metals ($\sim${}0.2 Solar), this column produces EW(O~{\sc vii}) $\approx$ 2$-$10 m\AA, depending on the projected radius. The column density is higher for galaxy groups, which are more massive and larger (cross section scales as R$^{2}_{200}$), but their space density is lower. The number of detectable lines per sight line depends on the space density of systems, and both simulations \citep{cen06, cen12} and simple estimates predict a detectable system every $\Delta$z $\approx$ 0.1$-$0.2 for EW $>$ 3 m\AA, and every $\Delta$z $\approx$ 0.06 for EW $>$ 1 m\AA.  This may be an underestimate for the detection rate, as recent observational work indicates a rate a few times higher \citep{nicastro18}.

To assess the detectability of such lines, we consider two telescope concepts: a MIDEX-class mission\footnote{Medium-class explorer, https://explorers.gsfc.nasa.gov/midex.html} with $R$=3000 and a collecting area of about 300 cm$^{{\rm 2}}$ \citep[such as \textit{Arcus};][]{smith17}, and a Large Strategic mission with $R$=6000 and a collecting area of about 4000 cm$^{{\rm 2}}$  \citep[\textit{Lynx};][]{gaskin18}). \textit{Arcus} could detect lines with EW$>$3~m\AA, while \textit{Lynx} could achieve the more ambitious goal of detecting absorption lines with EW $\lesssim$ 1 m\AA.

Either instrument could perform a blind survey up to a redshift of $z$=1.6 (for O~{\sc vii} He$\alpha$) and $z$=2.0 (for O~{\sc viii} Ly$\alpha$), with the limits set by severe absorption of X-rays by foreground Galactic gas below $E$=0.22~keV. Within this space, the next challenge is to identify absorber redshifts. Our calculations show that more than one absorption line will be detected for many systems, and we expect that identifications will be aided by UV absorption line studies of lower ionization lines (when available) and by optical redshift determinations of likely host systems. 

The $\sim$100 best background AGN targets have been defined \citep{breg15}, with a median AGN redshift is $z \sim 0.3$ (and a range up to $z<2$). Most absorption systems will occur with $z<0.6$.  Most of the absorption lines associated with galaxies will come from galaxies with L$_{{\rm opt}}$ $>$ 0.2L$^*$, based on models \citep{Qu2018b}. Such galaxies are brighter than $m_r = 22$~mag and are already detected by large surveys (e.g., Pan-STARRS), allowing spectroscopic follow-up.

\medskip
\noindent{\bf 2.2 Hosts of X-ray Absorption Systems: Galaxies, Galaxy Groups, and the Cosmic Web}  
\medskip

Once absorbers have been associated with hosts, it will be possible to study the properties of the gas in galaxy halos, in galaxy groups, and in the cosmic web. 

The observations will enable a measurement of the density profile up to (\textit{Arcus}) and beyond (\textit{Lynx}) R$_{200}$ (Fig.~3). Currently, the Milky Way is the only L* system studied in X-ray absorption, and the absorption is dominated by gas within 50~kpc \citep{breg18}. More massive galaxies, both early- and late-type, have X-ray emitting gas that is also typically seen within 50~kpc. These data can be described by a density profile $n \propto r^{{\rm -3/2}}$, but its behavior at larger radii is unclear. The blind survey will distinguish between competing models ($r^{-3/2}$, flattened, or NFW profiles) and determine whether the halo terminates at the splashback radius (Fig.~3). In combination with the metallicity, this density profile will robustly constrain the mass. 


In larger systems, such as galaxy groups, it will be possible to measure velocity information from the line profiles and determine the radial (infall) and azimuthal (rotation) velocity of the hot gas, as well as constrain the turbulent velocity from the line width. These will directly test galaxy formation models and enable a search for hot outflows. One generic prediction is that the hot gas is volume-filling, so the line profiles are likely different than the discrete, multi-component systems of UV-traced CGM clouds with a small filling factor \citep{stocke13, werk14}. Nevertheless, with $R>3000$ these measurements are possible \citep[e.g.,][]{miller16a}. Both {\it Arcus \/}and {\it Lynx \/}will measure velocities from centroids, but the higher spectral resolution of {\it Lynx \/}will resolve the lines into multiple components.

At the lowest EW, we anticipate detection of hot filaments in the Cosmic Web not associated with any particular galaxy or group. The filaments are not fully virialized, so the lines will be weaker than in galaxy halos, and so {\it Lynx\/}, with its greater sensitivity, is best suited for making a breakthrough. Such detections will be aided by redshifts of galaxies along the line of sight to detect candidate filaments and rule out absorption at the outskirts of (virialized) galaxy groups.
\smallskip

\begin{figure}
\vspace{-1.15in}
\hspace*{-0.95in}
\begin{minipage}{1.28\textwidth}
  \includegraphics[width=1.0\textwidth]{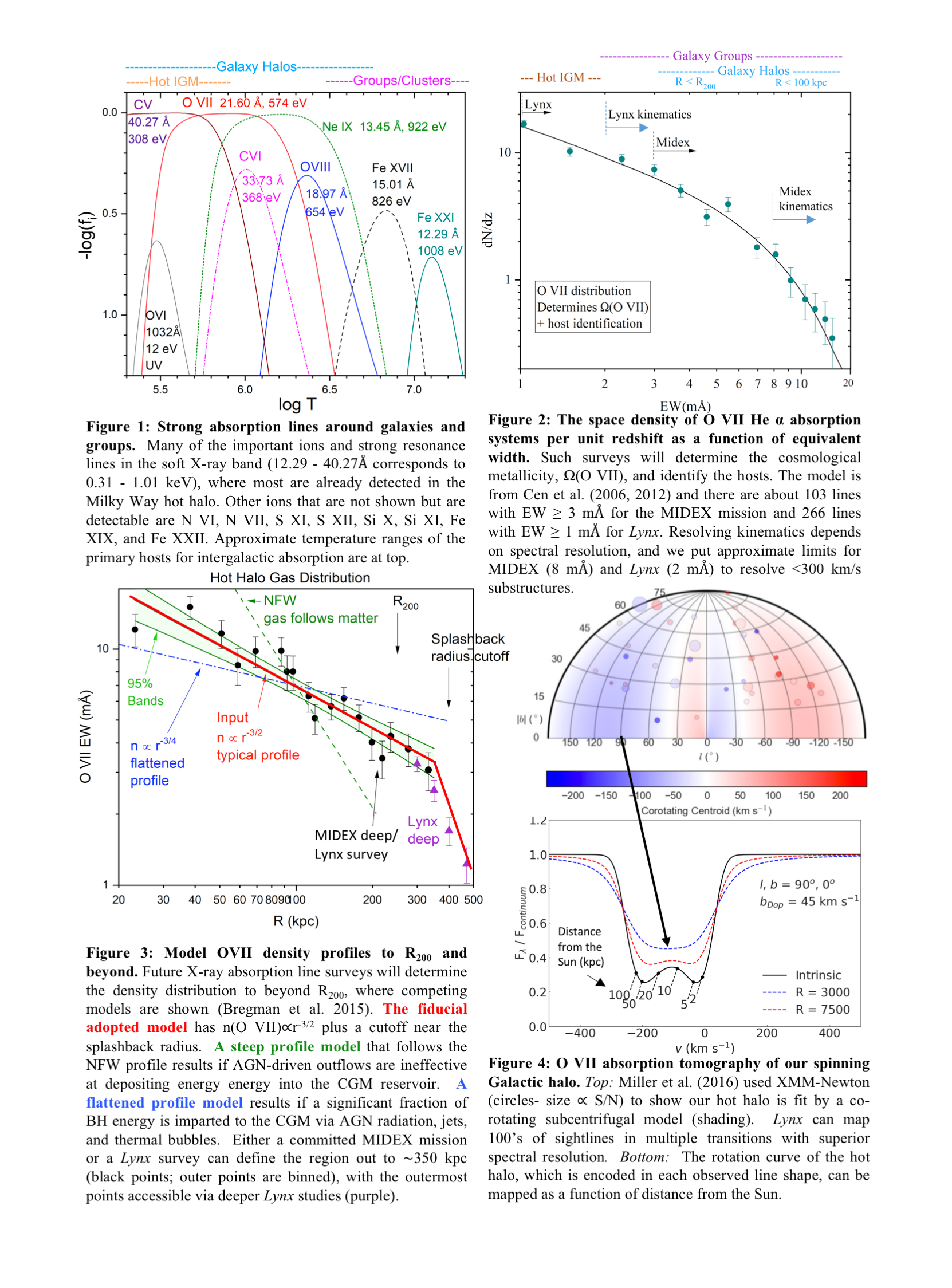}
\end{minipage}
\end{figure}

\medskip
\noindent{\bf \large 3. The Hot Circumgalactic Medium of the Milky Way}   \medskip

As in many fields, Milky Way studies of the hot CGM have led to important insights. The objectives are to measure the total gas mass in the Milky Way's hot halo and resolve Galactic O~{\sc vii} into components from (a) rotation and inflow, (b) turbulence, and (c) thermal broadening. Deriving the hot halo's mass and density profile requires accurate column density measurements with optical-depth corrections ($\tau$ = 0.5$-$3 at line center) obtained by measuring the line shapes and, independently, by measuring multiple lines of O~{\sc vii}, O~{\sc viii}, C~{\sc vi}, etc. 

This set of columns is fit with a density model to yield the mass of hot halo gas \citep{yli17,breg18}. The current uncertainty on the hot gas mass within R$_{{\rm 200}}$ is $\sim${}200\% ($<$ 8$\times$10$^{{\rm 10}}$ M$_{{\rm \odot}{}}$ within 250 kpc), but there is also a question regarding the proper density profile to use \citep{miller15,nicastro16,Nakashima2018}. The larger number of more accurate EWs from the two fiducial missions will reduce the statistical uncertainty to 25\% and resolve the debate over the shape of the halo. 

The improvement in the mass uncertainty results from measuring the column density beyond 50~kpc, as the mass to R$_{{\rm 200}}$ is obtained by
extrapolating the fit from R$<$50~kpc. The simplest approach is to measure the hot gas column to LMC/SMC targets and compare it to the column densities from sight lines passing through the entire Galactic halo \citep{breg18}. These same observations will provide a direct measure of the halo metallicity from the ratio of the O~{\sc vii} and O~{\sc viii} absorption EWs to their \textit{emission} measures (e.g., from \textit{Suzaku} or \textit{XMM-Newton}).

X-ray observations have shown that the Milky Way hot halo is rotating (within 50 kpc) and contains a significant amount of the total angular momentum \citep{hodges16}. This discovery provides critical new insights into galaxy accretion and formation \citep{opp18b}, but uncertainties are large ($v_{\rm rot}$= 183$\pm$41 km s$^{{\rm -}{1}}$) and it is unclear if the rotation axis is aligned with the MW disk.  New observations with high-resolution gratings will not only greatly reduce the mean uncertainty, it
will provide rotation information as a function of radius --{} a hot halo rotation curve \citep[][and Fig. 4]{miller16a}.  This requires
an instrument that can resolve absorption lines that are broadened by 200 km s$^{{\rm -1}}$ in key directions
({\it l\/} near 90$^{\circ}$, 270$^{\circ}$).  

The net inflow or outflow of hot CGM material is already limited to 5 M$_{{\rm \odot}{}}$ yr$^{{\rm -1}}$ from the observations with {\it XMM-Newton\/} at high Galactic latitudes \citep{hodges16}. Some models predict accretion values of 1-2  M$_{{\rm \odot}{}}$ yr$^{{\rm -1}}$ in order to replenish disk gas, and {\it Arcus \/}will determine the net accretion or outflow rate to levels of 2 M$_{{\rm \odot}{}}$ yr$^{{\rm -1}}$ (3$\sigma$), while {\it Lynx \/}will be a factor of two better \citep{miller16a}. The data would reveal if outflows occur in some directions (e.g., along the pole) concurrently with accretion onto the outer disk, providing vital insights into the growth and evolution of galaxies.

These same spectra will also yield the degree of turbulent line broadening along every sight line, revealing the magnitude and location of feedback. For example, feedback near the Galactic center is likely dominated by the enormous Fermi Bubbles, which shock the hot halo. Either mission will measure the expansion velocity of the shocked gas to a precision of $<$10\% through O~{\sc viii} absorption. A comparison of the expansion velocities to models will differentiate between an impulsive AGN outburst from Sgr~A* and ongoing star formation. 
\medskip

\noindent{\bf \large 4.  Recommendation for the Future of High Resolution X-Ray Spectroscopy}  \medskip

{\it Arcus\/} and {\it Lynx \/}will improve soft X-ray spectroscopy by 50 and 1500 times relative to current instruments (figure of merit is $R \times A_{\rm eff}$) respectively.  That will expand the number of observable X-ray targets from a few hundred to at least 10$^{{\rm 5}}$-10$^{{\rm 7}}$, assuming a constant space density.  This enormous expansion of parameter space will lead X-ray spectroscopy from being a niche field to a standard tool that touches every field of astrophysics. Here we have outlined how it will reveal the properties of the CGM and intergalactic medium, uncovering vital clues for the evolution of galaxies.  This undertaking cannot be accomplished with a single X-ray spectrometer, anymore than a single optical spectrometer on a single telescope would have been adequate for ground-based astronomy.  Grating spectrometers should be planned for most future X-ray missions, just as optical-IR spectrometers are present on
nearly every ground-based telescope.

\clearpage
\setlength{\bibsep}{0pt plus 0.3ex} 

\bibliographystyle{apj} 
\bibliography{WhitePaper_IGM_WHIM_v2_nofigs}

\end{document}